# Soft Computing Models for Network Intrusion Detection Systems

Ajith Abraham[1] and Ravi. Jain[2]

[1]Department of Computer Science, Oklahoma State University, USA
ajith.abraham@ieee.org
[2]School of Information Science, University of South Australia, Australia
ravi.jain@unisa.edu.au

**Abstract**: Security of computers and the networks that connect them is increasingly becoming of great significance. Computer security is defined as the protection of computing systems against threats to confidentiality, integrity, and availability. There are two types of intruders: external intruders, who are unauthorized users of the machines they attack, and internal intruders, who have permission to access the system with some restrictions. This chapter presents a soft computing approach to detect intrusions in a network. Among the several soft computing paradigms, we investigated fuzzy rule-based classifiers, decision trees, support vector machines, linear genetic programming and an ensemble method to model fast and efficient intrusion detection systems. Empirical results clearly show that soft computing approach could play a major role for intrusion detection.

**Keywords**: intrusion detection information security, soft computing

## 1 Introduction

The traditional prevention techniques such as user authentication, data encryption, the avoidance of programming errors and firewalls are used as the first line of defense for computer security. If a password is weak and is compromised, user authentication cannot prevent unauthorized use. Firewalls are vulnerable to errors in configuration and ambiguous or undefined security policies. They are generally unable to protect against malicious mobile code, insider attacks and unsecured modems. Programming errors cannot be avoided as the complexity of the system and application software is changing rapidly, leaving behind some exploitable weaknesses. Intrusion detection is therefore required as an additional wall for protecting systems [5][9]. Intrusion detection is useful not only in detecting successful intrusions, but also provides important information for timely countermeasures [11][13]. An intrusion is defined as any set of actions that attempt to compromise the integrity, confidentiality or availability of a resource. An attacker can gain access because of an error in the configuration of a system. In some cases it is possible to fool a system into giving access by misrepresenting oneself. An example is sending a TCP packet that has a forged source address that makes the packet appear to come from a trusted host. Intrusions may be classified into several types [12].

- Attempted break-ins, which are detected by typical behavior profiles or violations of security constraints.
- Masquerade attacks, which are detected by atypical behavior profiles or violations of security constraints.
- Penetration of the security control system, which are detected by monitoring for specific patterns of activity.
- Leakage, which is detected by atypical use of system resources.
- Denial of service, which is detected by atypical use of system resources.
- Malicious use, which is detected by atypical behavior profiles, violations of security constraints, or use of special privileges.

The process of monitoring the events occurring in a computer system or network and analyzing them for sign of intrusions is known as intrusion detection. Intrusion detection is classified into two types: misuse intrusion detection and anomaly intrusion detection.

Misuse intrusion detection uses well-defined patterns of the attack that exploit weaknesses in system and application software to identify the intrusions. These patterns are encoded in advance and used to match against the user behavior to detect intrusion.

Anomaly intrusion detection uses the normal usage behavior patterns to identify the intrusion. The normal usage patterns are constructed from the statistical measures of the system features, for example, the CPU and I/O activities by a particular user or program. The behavior of the user is observed and any deviation from the constructed normal behavior is detected as intrusion.

We have two options to secure the system completely, either prevent the threats and vulnerabilities which come from flaws in the operating system, as well as in the application programs, or detect them and take some action to prevent them in future and also repair the damage. It is impossible in practice, and even if possible, extremely difficult and expensive, to write a completely secure system. Transition to such a system for use in the entire world would be an equally difficult task. Cryptographic methods can be compromised if passwords and keys are stolen. No matter how secure a system is, it is vulnerable to insiders who abuse their privileges. There is an inverse relationship between the level of access control and efficiency. More access controls make a system less user-friendly and more likely to not be used.

An Intrusion Detection System (IDS) is a program that analyzes what happens or has happened during an execution and tries to find indications that the computer has been misused. An intrusion detection system does not eliminate the use of preventive mechanism but it works as the last defensive mechanism in securing the system. Data mining approaches are a relatively new technique for intrusion detection.

## 2 Intrusion Detection – A Data Mining Approach

Data mining is a relatively new approach for intrusion detection. Data mining approaches for intrusion detection were first implemented in mining audit data for automated models for intrusion detection [2][8]. The raw data is first converted into ASCII network packet information, which in turn is converted into connection level information. These connection level records contain connection features like service, duration, etc. Data mining algorithms are applied to this data to create models to detect intrusions. Data mining algorithms used in this approach are RIPPER (rule based classification algorithm), meta-classifier, frequent episode algorithm and association rules. These algorithms are applied to audit data to compute models that accurately capture the actual behavior of intrusions as well as normal activities.

The RIPPER algorithm was used to learn the classification model in order to identify normal and abnormal behavior [4]. Frequent episode algorithm and association rules together are used to construct frequent patterns from audit data records. These frequent patterns represent the statistical summaries of network and system activity by measuring the correlations among system features and the sequential co-occurrence of events. From the constructed frequent patterns the consistent patterns of normal activities and the unique intrusion patterns are identified and analyzed, and then used to construct additional features. These additional features are useful in learning the detection model more efficiently in order to detect intrusions. The RIPPER classification algorithm is then used to learn the detection model. A Meta classifier is used to learn the correlation of intrusion evidence from multiple detection models and to produce a combined detection model. The main advantage of this system is the automation of data analysis through data mining, which enables it to learn rules inductively, replacing manual encoding of intrusion patterns. However, some novel attacks may not be detected.

Audit data analysis and mining combine's association rules and classification algorithm to discover attacks in audit data [1]. Association rules are used to gather necessary knowledge about the nature of the audit data as the information about patterns within individual records can improve the classification efficiency. This system has two phases: training and detection. In the training phase a database of frequent item sets is created for the attack-free items by using only the attack-free data set. This serves as a profile against which frequent item sets found later will be compared. Next a sliding-window, on-line algorithm is used to find frequent item sets in the last $D$ connections and compares them with those stored in the attack-free database, discarding those that are deemed normal. In this phase a classifier is also trained to learn the model to detect the attack. In the detection phase a dynamic algorithm is used to produce item sets that are considered as suspicious and used by the classification algorithm already learned to classify the item set as attack, false alarm (normal event) or as unknown. Unknown attacks are the ones which are not able to detect either as false alarms or as known attacks. This method attempts to detect only anomaly attacks.

## 3 Soft Computing Models

Soft Computing (SC) is an innovative approach to construct computationally intelligent systems consisting of artificial neural networks, fuzzy inference systems, approximate reasoning and derivative free optimization methods such as evolutionary computation etc. In contrast with conventional artificial intelligence techniques which only deal with precision, certainty and rigor the guiding principle of soft computing is to exploit the tolerance for imprecision, uncertainty, low solution cost, robustness, partial truth to achieve tractability and better rapport with reality [15].

### 3.1 Fuzzy Rule Based Systems

Fuzzy logic has proved to be a powerful tool for decision making to handle and manipulate imprecise and noisy data. The notion central to fuzzy systems is that truth values (in fuzzy logic) or membership values (in fuzzy sets) are indicated by a value on the range [0.0, 1.0], with 0.0 representing absolute falseness and 1.0 representing absolute truth. A fuzzy system is characterized by a set of linguistic statements based on expert knowledge. The expert knowledge is usually in the form of *if-then* rules.

*Definition 1*: Let $X$ be some set of objects, with elements noted as $x$. Thus, $X = \{x\}$.

*Definition 2*: A fuzzy set $A$ in X is characterized by a membership function which are easily implemented by fuzzy conditional statements. In the case of fuzzy statement if the antecedent is true to some degree of membership then the consequent is also true to that same degree.

A simple rule structure: **If** antecedent **then** consequent

A simple rule: **If** variable$_1$ is low and variable$_2$ is high **then** output is benign **else** output is malignant

In a fuzzy classification system, a case or an object can be classified by applying a set of fuzzy rules based on the linguistic values of its attributes. Every rule has a weight, which is a number between 0 and 1 and this is applied to the number given by the antecedent. It involves 2 distinct parts. First the antecedent is evaluated, which in turn involves fuzzifying the input and applying any necessary fuzzy operators and second applying that result to the consequent known as inference. To build a fuzzy classification system, the most difficult task is to find a set of fuzzy rules pertaining to the specific classification problem.

We explored three fuzzy rule generation methods for intrusion detection systems. Let us assume that we have a $n$ dimensional c-class pattern classification problem whose pattern space is an $n$-dimensional unit cube $[0, 1]^n$. We also assume that $m$ patterns $x_p = (x_{p1},\ldots,x_{pn})$, $p = 1,2,\ldots,m,$ are given for generating fuzzy *if-then* rules where $x_p \in [0,1]$ for $p = 1,2,\ldots, m, i = 1,2,\ldots,n$ where $x_p \in [0,1]$ for $p = 1,2,\ldots, m, i = 1,2,\ldots,n$.

### *Rule Generation Based on the Histogram of Attribute Values (FR$_1$)*

In this method, use of histogram itself is an antecedent membership function. Each attribute is partitioned into 20 membership functions $f_h(.)$,

$h=1,2,\ldots,20$. The smoothed histogram $m_i^k(x_i)$ of class $k$ patterns for the $i^{th}$ attribute is calculated using the 20 membership functions $f_h(.)$ as follows:

$$m_i^k(x_i) = \frac{1}{m^k} \sum_{x_p \in Class\ k} f_h(x_{pi}) \qquad (1)$$

$$\text{for } \beta_{h-1} \leq x_i \leq \beta_h,\ h=1,2,\ldots,20$$

where $m_k$ is the number of Class $k$ patterns, $[\beta_{h-1},\beta_h]$ is the $h^{th}$ crisp interval corresponding to the 0.5-level set of the membership function $f_h(.)$

$$\beta_1 = 0,\ \beta_{20}=1, \qquad (2)$$

$$\beta_h = \frac{1}{20-1}\left(h-\frac{1}{2}\right) \text{ for } h=1,2,\ldots,19 \qquad (3)$$

The smoothed histogram in (1) is normalized so that its maximum value is 1. A single fuzzy *if-then* rule is generated for each class. The fuzzy if-then rule for the $k^{th}$ class can be written as

If $x_1$ is $A_1^k$ and ... and $x_n$ is $A_1^k$ then class $k$, \qquad (4)

where $A_i^k$ is an antecedent fuzzy set for the $i^{th}$ attribute. The membership function of $A_i^k$ is specified as

$$A_i^k(x_i) = \exp\left(-\frac{(x_i - \mu_i^k)^2}{2(\sigma_i^k)^2}\right) \qquad (5)$$

where $\mu_i^k$ is the mean of the $i^{th}$ attribute values $x_{pi}$ of class $k$ patterns, and $\sigma_i^k$ is the standard deviation. Fuzzy *if-then* rules for the two-dimensional two class pattern classification problem are written as follows:

If $x_3$ is $A_3^1$ and $x_4$ is $A_4^1$ *then* class 2 \qquad (6)

If $x_3$ is $A_3^2$ and $x_4$ is $\sqrt{a^2+b^2}$ *then* class 3 \qquad (7)

membership function of each antecedent fuzzy set is specified by the mean and the standard deviation of attribute values. For a new pattern $x_p = (x_{p3}, x_{p4})$, the winner rule is determined as follows:

$$A_3^*(x_{p3}) \cdot A_2^*(x_{p4}) = \max\left\{A_1^k(x_{p3}) \cdot A_2^k(x_{p4}) | k=1,2\right\} \qquad (8)$$

### Rule Generation Based on Partition of Overlapping Areas (FR$_2$)

Figure 1 demonstrates a simple fuzzy partition, where the two-dimensional pattern space is partitioned into 25 fuzzy subspaces by five fuzzy sets for each attribute (*S*: small, *MS*: medium small, *M*: medium, *ML*: medium large, *L*: large). A single fuzzy *if-then* rule is generated for each fuzzy subspace. Thus the number of possible fuzzy *if-then* rules in Figure 1 is 25.

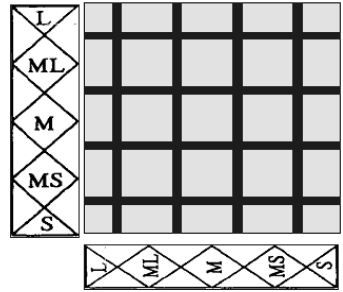

**Fig.1:** An example of fuzzy partition

One disadvantage of this approach is that the number of possible fuzzy *if-then* rules exponentially increases with the dimensionality of the pattern space. Because the specification of each membership function does not depend on any information about training patterns, this approach uses fuzzy if-then rules with certainty grades. The local information about training patterns in the corresponding fuzzy subspace is used for determining the consequent class and the grade of certainty. In this approach, fuzzy if-then rules of the following type are used:

*If* x$_1$ is $A_{j1}$ and ... and x$_n$ is $A_{jn}$ *then* class C$_j$,

with $CF = CF_j$, $j = 1, 2, ..., N$        (9)

where *j* indexes the number of rules, *N* is the total number of rules, $A_{ji}$ is the antecedent fuzzy set of the $i^{th}$ rule for the $i^{th}$ attribute, $C_j$ is the consequent class, and *CFj* is the grade of certainty. The consequent class and the grade of certainty of each rule are determined by the following simple heuristic procedure:

**Step 1:** Calculate the compatibility of each training pattern $x_p$ =($x_{p1}, x_{p2}, ..., x_{pn}$) with the $j^{th}$ fuzzy *if-then* rule by the following product operation:

$$\pi_j(x_p) = A_{j1}(x_{p1}) \times ... \times A_{jn}(x_{pn}), p = 1, 2, ..., m.$$   (10)

**Step 2:** For each class, calculate the sum of the compatibility grades of the training patterns with the $j^{th}$ fuzzy *if-then* rule $R_j$:

$$\beta_{class\ k}(R_j) = \sum_{x_p \in class\ k}^{n} \pi(x_p), \ k=1,2,...,c \qquad (11)$$

where $\beta_{class\ k}(R_j)$ the sum of the compatibility grades of the training patterns in class $k$ with the $j^{th}$ fuzzy if-then rule $R_j$.

**Step 3:** Find Class $A_j^*$ that has the maximum value $\beta_{class\ k}(R_j)$:

$$\beta_{class\ k_j^*} = Max\{\beta_{class\ 1}(R_j),...,\beta_{class\ c}(R_j)\} \qquad (12)$$

If two or more classes take the maximum value or no training pattern compatible with the $j^{th}$ fuzzy *if-then* rule (i.e., if $\beta_{Class\ k}(R_j)=0$ for $k=1,2,...,c$), the consequent class $C_i$ can not be determined uniquely. In this case, let $C_i$ be $\phi$.

**Step 4:** If the consequent class $C_i$ is 0, let the grade of certainty $CF_j$ be $CF_j = 0$. Otherwise the grade of certainty $CF_j$ is determined as follows:

$$CF_j = \frac{(\beta_{class\ k_j^*}(R_j) - \bar{\beta})}{\sum_{k=1}^{c} \beta_{class\ k}(R_j)} \qquad (13)$$

where $\bar{\beta} = \sum_{\substack{k=1 \\ k \neq k_j^*}} \frac{\beta_{Class\ k}(R_j)}{(c-1)}$

The above approach could be modified by partitioning only the overlapping areas as illustrated in Figure 2.

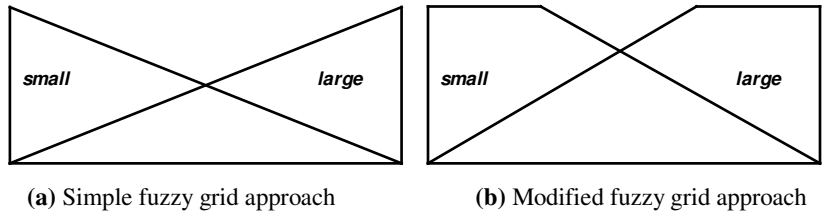

    **(a)** Simple fuzzy grid approach       **(b)** Modified fuzzy grid approach

**Fig. 2.** Fuzzy partition of each attribute

This approach generates fuzzy *if-then* rules in the same manner as the simple fuzzy grid approach except for the specification of each membership function. Because this approach utilizes the information about training patterns for specifying each membership function as mentioned in Section 2.1.1, the performance of generated fuzzy *if- then* rules is good even when we do not use the certainty grade of each rule in the classification phase. In this approach, the effect of introducing the certainty grade to each rule is not so important when compared to conventional grid partitioning.

### *Neural Learning of Fuzzy Rules (FR$_3$)*

The derivation of *if-then* rules and corresponding membership functions depends heavily on the *a priori* knowledge about the system under consideration. However there is no systematic way to transform experiences of knowledge of human experts to the knowledge base of a Fuzzy Inference System (FIS). In a fused neuro-fuzzy architecture, neural network learning algorithms are used to determine the parameters of fuzzy inference system (membership functions and number of rules). Fused neuro-fuzzy systems share data structures and knowledge representations. A common way to apply a learning algorithm to a fuzzy system is to represent it in a special neural network-like architecture. An Evolving Fuzzy Neural Network (EFuNN) implements a Mamdani type FIS and all nodes are created during learning. The nodes representing membership functions (MF) can be modified during learning. Each input variable is represented here by a group of spatially arranged neurons to represent a fuzzy quantization of this variable. New neurons can evolve in this layer if, for a given input vector, the corresponding variable value does not belong to any of the existing MF to a degree greater than a membership threshold. Technical details of the learning algorithm are given in [16].

### 3.2. Linear Genetic Programming (LGP)

Linear genetic programming is a variant of the GP technique that acts on linear genomes [3]. Its main characteristics in comparison to tree-based GP are that the evolvable units are not expressions of a functional programming language (like LISP), but the programs of an imperative language (like c/c ++). An alternate approach is to evolve a computer program at the machine code level, using lower level representations for the individuals. This can tremendously hasten the evolution process as, no matter how an individual is initially represented, finally it always has to be represented as a piece of machine code, as fitness evaluation requires physical execution of the individuals.

The basic unit of evolution here is a native machine code instruction that runs on the floating-point processor unit (FPU). Since different instructions may have different sizes, here instructions are clubbed up together to form instruction blocks of 32 bits each. The instruction blocks hold one or more native machine code instructions, depending on the sizes of the instructions. A crossover point can occur only between instructions and is prohibited from occurring within an instruction. However the mutation operation does not have any such restriction. In this research a steady state genetic programming approach was used to manage the memory more effectively [1].

### 3.3. Decision Trees (DT)

Intrusion detection can be considered as classification problem where each connection or user is identified either as one of the attack types or normal based on some existing data. Decision trees work well with large data sets. This is important as large amounts of data flow across computer networks. The high performance of decision trees makes them useful in real-time intrusion detection. Decision trees construct easily interpretable models, which is useful for a security officer to inspect and edit. These models can also be used in the rule-based models with minimum processing [7]. Generalization accuracy of decision trees is another useful property for intrusion detection model. There will always be new attacks on the system, which are small variations of known attacks after the intrusion detection models are built. The ability to detect these new intrusions is possible due to the generalization accuracy of decision trees.

### 3.4. Support Vector Machines (SVM)

Support Vector Machines have been proposed as a novel technique for intrusion detection. SVM maps input (real-valued) feature vectors into a higher dimensional feature space through some nonlinear mapping. SVMs are powerful tools for providing solutions to classification, regression and density estimation problems. These are developed on the principle of structural risk minimization. Structural risk minimization seeks to find a hypothesis for which one can find the lowest probability of error. The structural risk minimization can be achieved by finding the hyper plane with maximum separable margin for the data [14]. Computing the hyper plane to separate the data points, i.e. training a SVM, leads to a quadratic optimization problem. SVM uses a feature called a kernel to solve this problem. A kernel transforms linear algorithms into nonlinear ones via a map into feature spaces. SVMs classify data by using these support vectors, which are members of the set of training inputs that outline a hyper plane in feature space.

## 4.0 Attribute Deduction in Intrusion Detection Systems

Since the amount of audit data that an IDS needs to examine is very large even for a small network, analysis is difficult even with computer assistance because extraneous features can make it harder to detect suspicious behavior patterns. Complex relationships exist between features, which are difficult for humans to discover. IDS must therefore reduce the amount of data to be processed. This is very important if real-time detection is desired. The easiest way to do this is by doing an intelligent input feature selection. Certain features may contain false correlations, which hinder the process of detecting intrusions. Further, some features may be redundant since the information they add is contained in other features. Extra features can increase computation time, and can impact the accuracy of IDS. Feature selection improves classification by searching for the subset of features, which best classifies the training data.

Feature selection is done based on the contribution the input variables made to the construction of the decision tree. Feature importance is determined by the role of each input variable either as a main splitter or as a surrogate. Surrogate splitters are defined as back-up rules that closely mimic the action of primary splitting rules. Suppose that, in a given model, the algorithm splits data according to variable 'protocol_type' and if a value for 'protocol_type' is not available, the algorithm might substitute 'flag' as a good surrogate. Variable importance, for a particular variable is the sum across all nodes in the tree of the improvement scores that the predictor has when it acts as a primary or surrogate (but not competitor) splitter. Example, for node $i$, if the predictor appears as the primary splitter then its contribution towards importance could be given as $i_{importance}$. But if the variable appears as the $n^{th}$ surrogate instead of the primary variable, then the importance becomes $i_{importance} = (p^n) * i_{improvement}$ in which $p$ is the 'surrogate improvement weight' which is a user controlled parameter set between (0-1) [17].

## 5.0 Intrusion Detection Data

In 1998, DARPA intrusion detection evaluation program created an environment to acquire raw TCP/IP dump data for a network by simulating a typical U.S. Air Force LAN [10]. The LAN was operated like a real environment, but was blasted with multiple attacks. For each TCP/IP connection, 41 various quantitative and qualitative features were extracted.

Of these a subset of 494,021 data were used for our studies, of which 20% represent normal patterns [6]. Different categories of attacks are summarized in Figure 4. Attack types fall into four main categories:

### *DoS: Denial of Service*

Denial of Service (DoS) is a class of attack where an attacker makes a computing or memory resource too busy or too full to handle legitimate requests, thus denying legitimate users access to a machine. There are different ways to launch DoS attacks: by abusing a computer's legitimate features; by targeting the implementation bugs; or by exploiting a system's miss configurations. DoS attacks are classified based on the services that an attacker renders unavailable to legitimate users.

### *R2L: Unauthorized Access from a Remote Machine*

A remote to user (R2L) attack is a class of attack where an attacker sends packets to a machine over a network, then exploits the machine's vulnerability to illegally gain local access as a user. There are different types of R2U attacks; the most common attack in this class is done using social engineering.

### *U2Su: Unauthorized Access to Local Super User (root)*

User to root (U2Su) exploits are a class of attacks where an attacker starts out with access to a normal user account on the system and is able to exploit vulnerability to gain root access to the system. Most common exploits in this class of attacks are regular buffer overflows, which are caused by regular programming mistakes and environment assumptions.

### *Probing: Surveillance and Other Probing*

Probing is a class of attack where an attacker scans a network to gather information or find known vulnerabilities. An attacker with a map of machines and services that are available on a network can use the information to look for exploits. There are different types of probes: some of them abuse the computer's legitimate features; some of them use social engineering techniques. This class of attack is the most common and requires very little technical expertise.

## 6.0 Experiment Setup and Results

The data for our experiments was prepared by the 1998 DARPA intrusion detection evaluation program by MIT Lincoln Labs [10]. The data set contains 24 attack types that could be classified into four main categories namely *Denial of Service (DoS), Remote to User (R2L), User to Root (U2R)* and *Probing*. The original data contains 744 MB data with 4,940,000 records. The data set has 41 attributes for each connection record plus one class label. Some features are derived features, which are useful in distinguishing normal connection from attacks. These features are either continuous or discrete. Some features examine only the connections in the past two seconds that have the same destination host as the current connection, and calculate statistics related to protocol behavior, service, etc. These are called same host features. Some features examine only the connections in the past two seconds that have the same service as the current connection and are called same service features. Some other connection records were also sorted by destination host, and features were constructed using a window of 100 connections to the same host instead of a time window. These are called host-based traffic features. R2L and U2R attacks don't have any sequential patterns like DoS and Probe because the former attacks have the attacks embedded in the data packets whereas the later attacks have many connections in a short amount of time. So some features that look for suspicious behavior in the data packets like number of failed logins are constructed and these are called content features.

Our experiments have three phases namely data reduction, training phase and testing phase. In the data reduction phase, important variables for real-time intrusion detection are selected by feature selection. In the training phase, the different soft computing models were constructed using the training data to give maximum generalization accuracy on the unseen data. The test data is then passed through the saved trained model to detect intrusions in the testing phase. The 41 features are labeled as shown in Table 1 and the class label is named as *AP*. This data set has five different classes namely *Normal, DoS, R2L, U2R* and *Probes.* The training and test comprises of 5,092 and 6,890 records respectively [6].

Our initial research was to reduce the number of variables. Using all 41 variables could result in a big IDS model, which could be an overhead for online detection. The experiment system consists of two stages: Network training and performance evaluation. All the training data were scaled to (0-1). The decision tree approach described in Section 4 helped us to reduce the number of variables to 12 variables. The list of reduced variables is illustrated in Table2.

Table 1. Variables for intrusion detection data set

| Variable No. | Variable name | Variable type | Variable label |
|---|---|---|---|
| 1 | duration | continuous | A |
| 2 | protocol_type | discrete | B |
| 3 | service | discrete | C |
| 4 | flag | discrete | D |
| 5 | src_bytes | continuous | E |
| 6 | dst_bytes | continuous | F |
| 7 | land | discrete | G |
| 8 | wrong_fragment | continuous | H |
| 9 | urgent | continuous | I |
| 10 | hot | continuous | J |
| 11 | num_failed_logins | continuous | K |
| 12 | logged_in | discrete | L |
| 13 | num_compromised | continuous | M |
| 14 | root_shell | continuous | N |
| 15 | su_attempted | continuous | O |
| 16 | num_root | continuous | P |
| 17 | num_file_creations | continuous | Q |
| 18 | num_shells | continuous | R |
| 19 | num_access_files | continuous | S |
| 20 | num_outbound_cmds | continuous | T |
| 21 | is_host_login | discrete | U |
| 22 | is_guest_login | discrete | V |
| 23 | count | continuous | W |
| 24 | srv_count | continuous | X |
| 25 | serror_rate | continuous | Y |
| 26 | srv_serror_rate | continuous | X |
| 27 | rerror_rate | continuous | AA |
| 28 | srv_rerror_rate | continuous | AB |
| 29 | same_srv_rate | continuous | AC |
| 30 | diff_srv_rate | continuous | AD |
| 31 | srv_diff_host_rate | continuous | AE |
| 32 | dst_host_count | continuous | AF |
| 33 | dst_host_srv_count | continuous | AG |
| 34 | dst_host_same_srv_rate | continuous | AH |
| 35 | dst_host_diff_srv_rate | continuous | AI |
| 36 | dst_host_same_src_port_rate | continuous | AJ |
| 37 | dst_host_srv_diff_host_rate | continuous | AK |
| 38 | dst_host_serror_rate | continuous | AL |
| 39 | dst_host_srv_serror_rate | continuous | AM |
| 40 | dst_host_rerror_rate | continuous | AN |
| 41 | dst_host_srv_rerror_rate | continuous | AO |

Table 2. Reduced variable set

| C, E, F, L, W, X, Y, AB, AE, AF, AG, AI |
|---|

Using the original and reduced data sets, we performed a 5-class classification. The (training and testing) data set contains 11,982 randomly generated points from the data set representing the five classes, with the number of data from each class proportional to its size, except that the smallest class is completely included. The set of 5,092 training data and 6,890 testing data are divided in to five classes: normal, probe, denial of service attacks, user to super user and remote to local attacks. The datasets contain a total of 24 training attack types, with an additional 14 types in the test data only. Where the attack is a collection of different types of instances that belong to the four classes described earlier and the other is the normal data. The normal data belongs to class 1, probe belongs to class 2, denial of service belongs to class 3, user to super user belongs to class 4, remote to local belongs to class 5. All the IDS models are trained and tested with the same set of data.

We examined the performance of all three fuzzy rule based approaches ($FR_1$, $FR_2$ and $FR_3$) mentioned in Section 3.1. When an attack is correctly classified the grade of certainty is increased and when an attack is misclassified the grade of certainty is decreased. A learning procedure is used to determine the grade of certainty. Triangular membership functions were used for all the fuzzy rule based classifiers. We used 4 triangular membership functions for each input variable for the EFuNN training ($FR_3$). A sensitivity threshold Sthr = 0.95 and error threshold Errthr = 0.05 was used for all the classes. 89 rule nodes were developed during the one pass learning [17].

The settings of various linear genetic programming system parameters are of utmost importance for successful performance of the system. The population space has been subdivided into multiple subpopulation or demes. Migration of individuals among the subpopulations causes evolution of the entire population. It helps to maintain diversity in the population, as migration is restricted among the demes. Table 3 depicts the parameter settings used for LGP experiments. The tournament size was set at 120,000 for all the 5 classes. Figure 3 demonstrates the growth in program length during 120,000 tournaments and the average fitness values for detecting normal patterns (class 1). More illustrations are available in [1].

**Table 1.** Parameter settings for linear genetic programming

| Parameter | Normal | Probe | DoS | U2Su | R2L |
| --- | --- | --- | --- | --- | --- |
| Population size | 2048 | 2048 | 2048 | 2048 | 2048 |
| Tournament size | 8 | 8 | 8 | 8 | 8 |
| Mutation frequency (%) | 85 | 82 | 75 | 86 | 85 |
| Crossover frequency (%) | 75 | 70 | 65 | 75 | 70 |
| Number of demes | 10 | 10 | 10 | 10 | 10 |
| Maximum program size | 256 | 256 | 256 | 256 | 256 |

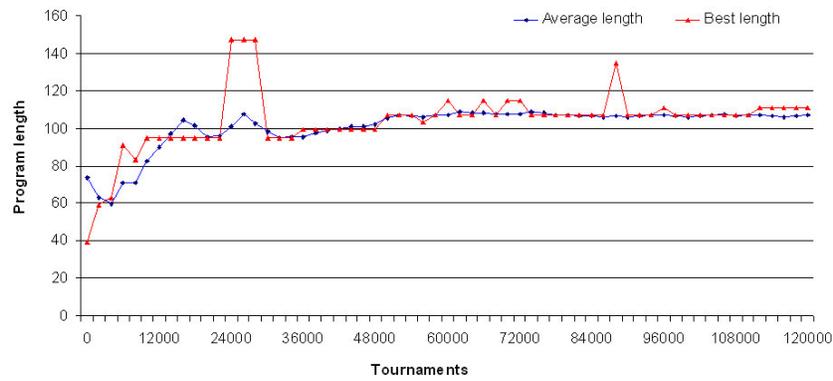

**(a)**

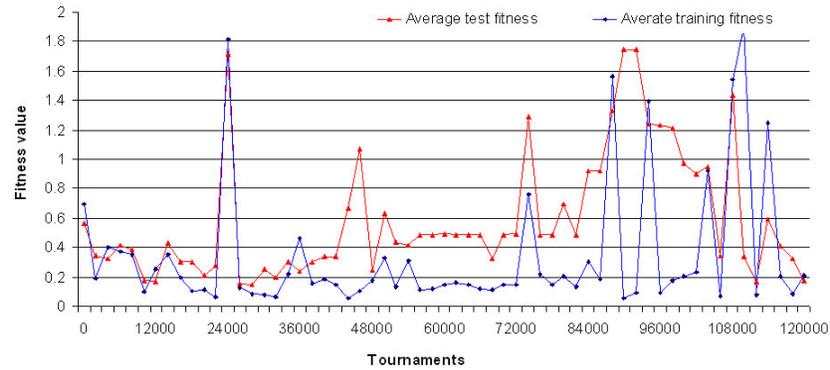

**(b)**

**Fig. 3.** LGP performance for the detection of normal patterns (a) growth in program length (b) average fitness

Our trial experiments with SVM revealed that the polynomial kernel option often performs well on most of the datasets. We also constructed decision trees using the training data and then testing data was passed through the constructed classifier to classify the attacks [12].

**Table 3.** Performance comparison using full data set

| Attack type | Classification accuracy on test data set (%) | | | | | |
|---|---|---|---|---|---|---|
| | $FR_1$ | $FR_2$ | $FR_3$ | DT | SVM | LGP |
| Normal | 40.44 | 100.00 | 98.26 | 99.64 | 99.64 | 99.73 |
| Probe  | 53.06 | 100.00 | 99.21 | 99.86 | 98.57 | 99.89 |
| DOS    | 60.99 | 100.00 | 98.18 | 96.83 | 99.92 | 99.95 |
| U2R    | 66.75 | 100.00 | 61.58 | 68.00 | 40.00 | 64.00 |
| R2L    | 61.10 | 100.00 | 95.46 | 84.19 | 33.92 | 99.47 |

**Table 4.** Performance comparison using reduced data set

| Attack type | Classification accuracy on test data set (%) | | | | | |
|---|---|---|---|---|---|---|
| | $FR_1$ | $FR_2$ | $FR_3$ | DT | SVM | LGP |
| Normal | 74.82 | 79.68 | 99.56 | 100.00 | 99.75 | 99.97 |
| Probe  | 45.36 | 89.84 | 99.88 | 97.71  | 98.20 | 99.93 |
| DOS    | 60.99 | 60.99 | 98.99 | 85.34  | 98.89 | 99.96 |
| U2R    | 94.11 | 99.64 | 65.00 | 64.00  | 59.00 | 68.26 |
| R2L    | 91.83 | 91.83 | 97.26 | 95.56  | 56.00 | 99.98 |

A number of observations and conclusions are drawn from the results illustrated in Tables 3 and 4. Using 41 attributes, the $FR_2$ method gave 100% accuracy for all the 5 classes, showing the importance of fuzzy inference systems. For the full data set, LGP outperformed decision trees and support vector machines in terms of detection accuracies (except for one class).

The reduced dataset seems to work very well for most of the classifiers except the fuzzy classifier ($FR_2$). For detecting U2R attacks $FR_2$ gave the best accuracy. Due to the tremendous reduction in the number of attributes (about 70% less), we are able to design a computational efficient intrusion detection system. Since a particular classifier could not provide accurate results for all the classes, we propose to use an ensemble approach as demonstrated in Figure 4. The proposed ensemble model could detect all the attacks with high accuracy (lowest accuracy being 99.64%) with only 12 input variables. Ensemble performance is summarized in Table 5.

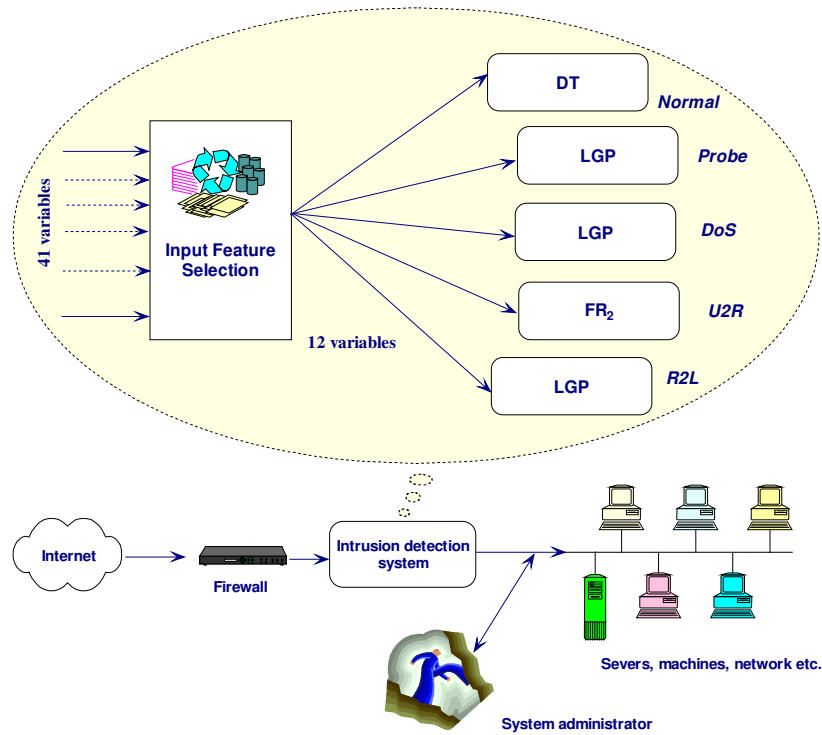

**Fig. 4.** IDS architecture using an ensemble of intelligent paradigms

**Table 5.** Performance of the ensemble method

| Attack type | Ensemble classification accuracy on test data (%) |
|---|---|
| Normal | 100.00 |
| Probe | 99.93 |
| DOS | 99.96 |
| U2R | 99.64 |
| R2L | 99.98 |

In some classes the accuracy figures tend to be very small and may not be statistically significant, especially in view of the fact that the 5 classes of patterns differ in their sizes tremendously. For example only 27 data sets were available for training the U2R class. More definitive conclusions can only be made after analyzing more comprehensive sets of network traffic.

## 7.0 Conclusions

In this chapter, we have illustrated the importance of soft computing paradigms for modeling intrusion detection systems. For real time intrusion detection systems, LGP would be the ideal candidate as it can be manipulated at the machine code level. Overall, the fuzzy classifier (FR$_2$) gave 100% accuracy for all attack types using all the 41 attributes. The proposed ensemble approach requires only 12 input variables. More data mining techniques are to be investigated for attribute reduction and enhance the performance of other soft computing paradigms.

## Acknowledgements

Authors would like to thank S. Chebrulu and S. Peddabachigari (Oklahoma State University, USA) for the various contributions during the different stages of this research.